\pdfoutput=1

\documentclass[
amsmath,
amssymb,
notitlepage,
aip,
jap,
citeautoscript,
reprint
]{revtex4-1}

\usepackage[greek, ngerman, english]{babel}
\usepackage{graphicx}
\usepackage{siunitx}
\usepackage{hyperref}
\usepackage{nicefrac}
\usepackage{color}

\DeclareSIUnit{\Electronmass}{\text{\ensuremath{m_{0}}}}

\begin{document}

\title{Charge transfer at organic-inorganic interfaces}
\date{\today}

\author{I.~Meyenburg}
\affiliation{Department of Physics and Material Sciences Center, Philipps-University Marburg, Renthof 5, 35032 Marburg, Germany}

\author{J.~Falgenhauer}
\affiliation{Institute of Applied Physics, Justus-Liebig-University Giessen, Heinrich-Buff-Ring 16, 35392 Giessen, Germany}

\author{N.W.~Rosemann}
\affiliation{Institute of Experimental Physics I, Justus-Liebig-University Giessen, Heinrich-Buff-Ring 16, 35392 Giessen, Germany}

\author{S.~Chatterjee}
\affiliation{Institute of Experimental Physics I, Justus-Liebig-University Giessen, Heinrich-Buff-Ring 16, 35392 Giessen, Germany}

\author{D.~Schlettwein}
\affiliation{Institute of Applied Physics, Justus-Liebig-University Giessen, Heinrich-Buff-Ring 16, 35392 Giessen, Germany}

\author{W.~Heimbrodt}
\email{Wolfram.Heimbrodt@physik.uni-marburg.de}
\affiliation{Department of Physics and Material Sciences Center, Philipps-University Marburg, Renthof 5, 35032 Marburg, Germany}

\begin{abstract}
We studied the electron transfer from excitons in adsorbed indoline dye layers across the organic-inorganic interface. The hybrids consist of indoline derivatives on the one hand and different inorganic substrates (TiO$_2$, ZnO, SiO$_2$(0001), fused silica) on the other.  
We reveal the electron transfer times from excitons in dye layers to the organic-inorganic interface by analyzing the photoluminescence transients of the dye layers after femtosecond excitation and applying kinetic model calculations.
A correlation between the transfer times and four parameters have been found:(i) the number of anchoring groups, (ii) the distance between the dye and the organic-inorganic interface, which was varied by the alkyl-chain lengths between the carboxylate anchoring group and the dye, (iii) the thickness of the adsorbed dye layer, and (iv) the level alignment between the excited dye ($\pi ^*$-level) and the conduction band minimum of the inorganic semiconductor.     

\end{abstract}

\maketitle

\begin{figure*}
\includegraphics[width=1\textwidth]{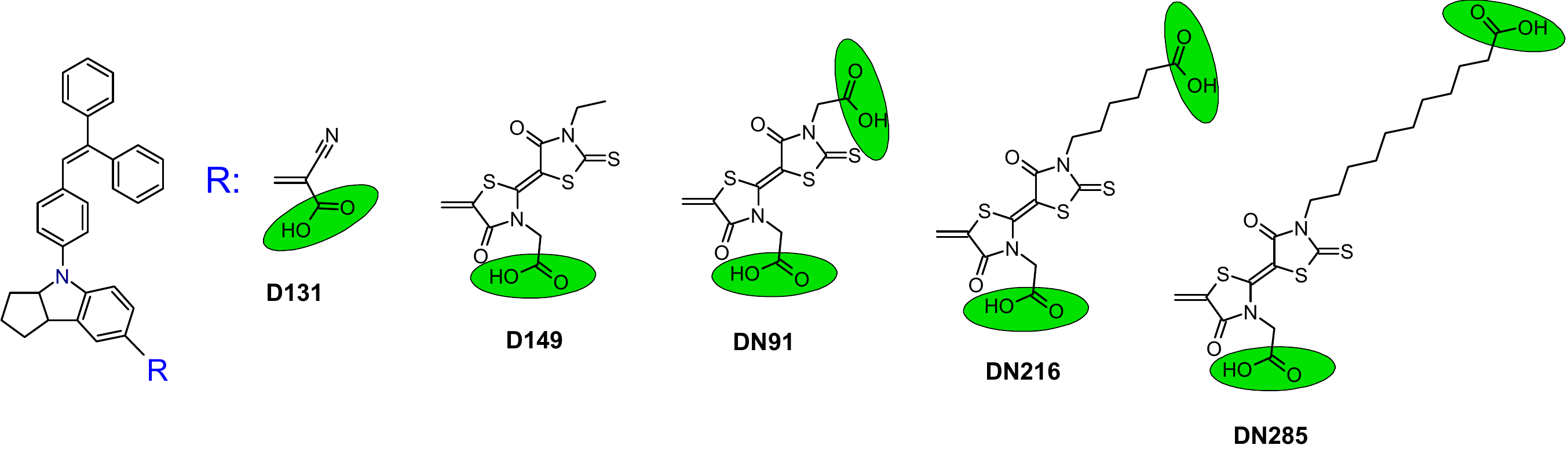}
\caption{Schematic picture of the chemical structure of the various indoline derivatives.}
\label{Fig_1}
\end{figure*}
\section{Introduction}
\label{sec:intro}

Organic-inorganic semiconductor hybrids are promising functional materials for advanced, optoelectronic devices. 
For example, dye-sensitized solar cells of the Gr\"atzel type \cite{Graetzel1991} are candidates for commercial photovoltaic applications.
Despite the relatively low quantum efficiencies compared to silicon based solar cells and to the recently reported perovskite based solar cells \cite{Liu2013}, there is continuous interest in Gr\"atzel type solar cells, as they are low cost devices and thus provide short energy payback times.

An important step to achieving higher quantum efficiencies in such dye-sensitized solar cells is an optimized electron transfer, i.e., the electron injection yield from the dye into the conduction band of the semiconductor at the organic-inorganic interface\cite{Listorti2011}. Physically, a mandatory prerequisite is the best possible electronic coupling between the excited dye sensitizers and the conduction band of the semiconductor. The relevant parameters for injection consist in (i) the level alignment between the excited state of the dye and the conduction band minimum (CBM) of the semiconductor as well as (ii) the mean distance between the excited sensitizer molecules and the semiconductor surface.

To date, various different semiconductor materials and dyes were suggested as efficient organic-inorganic hybrids for solar cell application (see, e.g., \cite{gratzel2005solar} and the review by Ooyama and Harima \cite{Ooyama2012}). 
Overall, TiO$_2$ is the prototypical n-type semiconductor. 
This material was already suggested in the original realization by O'Regan and Gr\"atzel \cite{Graetzel1991}. 
It is a wide bandgap material and therefore transparent for a sufficient spectral range of sunlight. 
A considered viable alternative is ZnO\cite{falgenhauer2012stable, rudolph2013influence,rohwer2013ultrafast, Rohwer2015}.
The location and type of anchoring group also plays a key role for the electronic coupling to the substrate \cite{Listorti2011, She2007}. 
The carboxylate anchoring group used in our studies have been shown to favor the electron injection into a metal oxide surface \cite{Listorti2011, DosSantos2011}.  

 Pump-probe transient absorption spectroscopy on the ps timescale is the most widely used method to study the excitation dynamics of dyes in solution \cite{fakis2013time, lohse2011ultrafast} and to determine electron transfer times \cite{Rohwer2015, Sobus2014, Fakis2011,oum2012ultrafast}. Such measurements typically use high-energy fs pulses for excitation. The depletion of the excited state is determined by the charge transfer (CT) but can be influenced also by the relaxation and formation of excitons as well as charge accumulation and respective screening effects. 
In this paper, we study the CT times for a variety of indoline dyes and semiconductor substrates by time-resolved photoluminescence (TRPL). 
In particular, we study the kinetics of the excitons in the dye layers after formation and relaxation from the primarily excited higher energy levels and determine the CT times from the exciton states to the interface at low excitation densities.  
We will show that the respective transfer times can differ substantially from times determined by pump-probe measurements. 
     
The excitonic PL transients are a complicated interplay of various processes influencing the transition probability. On one hand, the effective medium or molecule-molecule interaction changes the lifetime; on the other hand, many loss and energy transfer channels influence the lifetime. The resulting non-exponential curves are commonly fitted by a sum of several exponential functions or even by stretched exponentials, leaving the interpretation of the resulting transients as a sophisticated complex challenge. 
Here, we apply kinetic model calculations to describe the decay curves in order to reveal the charge transfer times from the excitons to the interface. Experimentally, we vary the substrate and hence the energy level alignment. Additionally we vary the geometric alignment of the dyes relative to the organic-inorganic interface and study the influence of different anchor groups.


\section{Experimental Details}
\label{sec:exp}

We study a series of five different indoline derivatives. 
All share the identical indoline chromophore, while the anchor groups are varied. 
The schematic structures are depicted in Fig.\ref{Fig_1}. D131 and D149 have only one carboxyl anchor group but differ in the distance between the carboxyl group and the indoline chromophore. All the other indoline derivatives have two carboxyl anchor groups and differ from each other in the length of the anchor chain of the second carboxyl group.

We prepared the sensitized mesoporous samples by immersing films of the mesoporous semiconductors into a solution of indoline dye and acetonitrile\,/\,tert.\,butanol (50\%/50\%). This "soaking" is the standard technique for Gr\"atzel type dye sensitized solar cells. The substrates for soaking are mesoporous ZnO (ZnO(meso)) or mesoporous anatase TiO$_2$ (TiO$_2$(meso)). TiO$_2$(meso) samples consist of a 10\,$\mu$m thick layer on fluorine-doped tin oxide (FTO) glass which is built up by $\sim $20\,nm thick particles of TiO$_2$. The layer is prepared by applying a TiO$_2$-gel and sintering at $450^\circ\text{C}$. Mesoporous ZnO was prepared by an electrochemical deposition of ZnO on FTO glass in the presence of a structure-directing agent, using KCl as electrolyte component. The deposited film was left overnight in an aqueous KOH solution to remove the structure-directing agent and dried before the sensitization (see \cite{richter2015influence} for experimental details). \\
Indoline dye layers were prepared by drop casting the same solution as used for soaking on flat substrates, i.e., on single crystals of ZnO(0001), rutile TiO$_2$(100), and SiO$_2$(0001) (Crystec) as well as on standard fused silica.
    
We performed photoluminescence (PL) and absorption measurements by means of a standard setup with a high-resolution grating spectrometer. For the steady-state PL, a laser diode of 442\,nm (2.8\,eV) and for the optical absorption measurements a white light lamp were used. \\

The time-resolved PL experiments were performed using a standard streak-camera setup. Details are given for example in\cite{arbiol2012self}. The 100-fs pulses of the tunable Ti:Sapphire laser were frequency doubled to 2.8\,eV (442\,nm) for most experiments; a 475\,nm color-glass edge filter was used in the detection to suppress scattered laser light. Sensitive measurements on thin layers of D149 and D131 samples were excited at 3.06\,eV (405\,nm) in combination with an edge filter at 2.76\,eV (450\,nm).


\section{Results and Discussion}
\label{sec:discussion}

\begin{figure}
\includegraphics[width=0.5\textwidth]{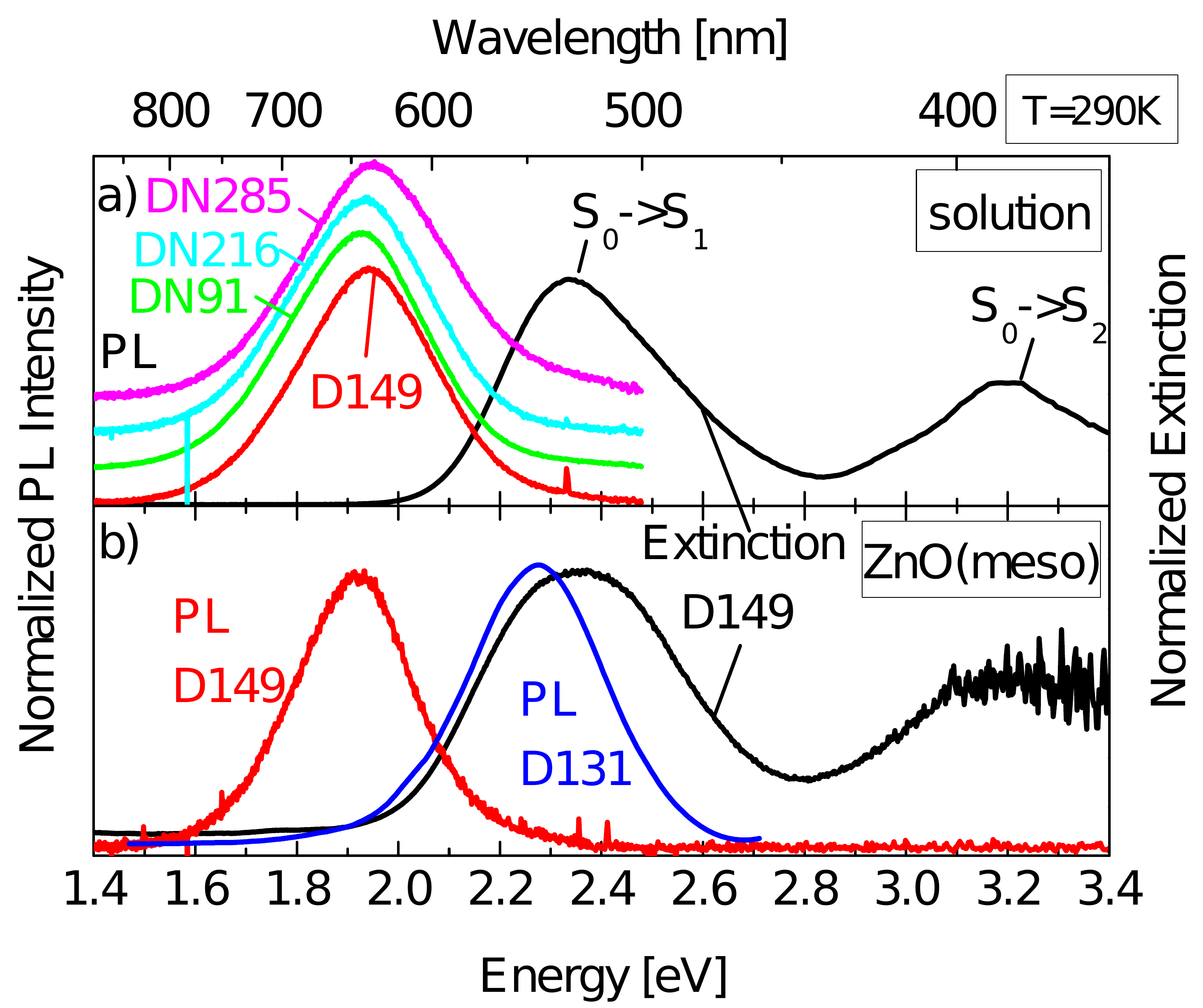}
\caption{(a) Photoluminescence of indoline derivatives and exctinction spectzra of D149 in solution at room temperature. (b) Photoluminescence of D149 and D131 and exctintion of D149 on mesoporous ZnO at room temperature.}
\label{Fig_2}
\end{figure}

The PL spectra of all indoline derivatives in solution are summarized in Fig.~\ref{Fig_2}a. 
The different lengths of the chains do not substantially influence the $\pi \to \pi ^\star$  transitions.\\ 
The $S_1 \to S_0$ PL transition energy is very similar for D149 and DN285. However, it exhibits a slight redshift for the shorter anchor chains in DN216 and DN91.\\ 
For comparison, a typical PL spectrum of a D149 layer deposited on ZnO(meso) is depicted in Fig.~\ref{Fig_2}b. 
Compared to Fig.~\ref{Fig_2}a, we find a typical red-shift of the emission due to molecule-molecule interactions. 
This red shift is observed for all different substrates (TiO$_2$(100), TiO$_2$(meso), ZnO(0001), ZnO(meso), fused silica), however, the substrates have no significant influence on the spectral width and shape. 
Comparing the exctinction spectra in Fig.~\ref{Fig_2} of D149 in solution and of a D149 layer on ZnO(meso) reveals a broadening due to the weak van-der-Waals molecule-molecule interaction and the respective band formation and dispersion. The different energies of the PL and extinction transitions are caused by the Stokes shift expected according to the Franck-Condon principle.\\ 
In Fig.~\ref{Fig_2}b the PL of a D131 layer on ZnO(meso) is depicted for comparison. Clearly, the PL is strongly blue shifted compared to D149; this is typically for the smaller $\pi$ electron system.

\begin{figure}
\includegraphics[width=0.5\textwidth]{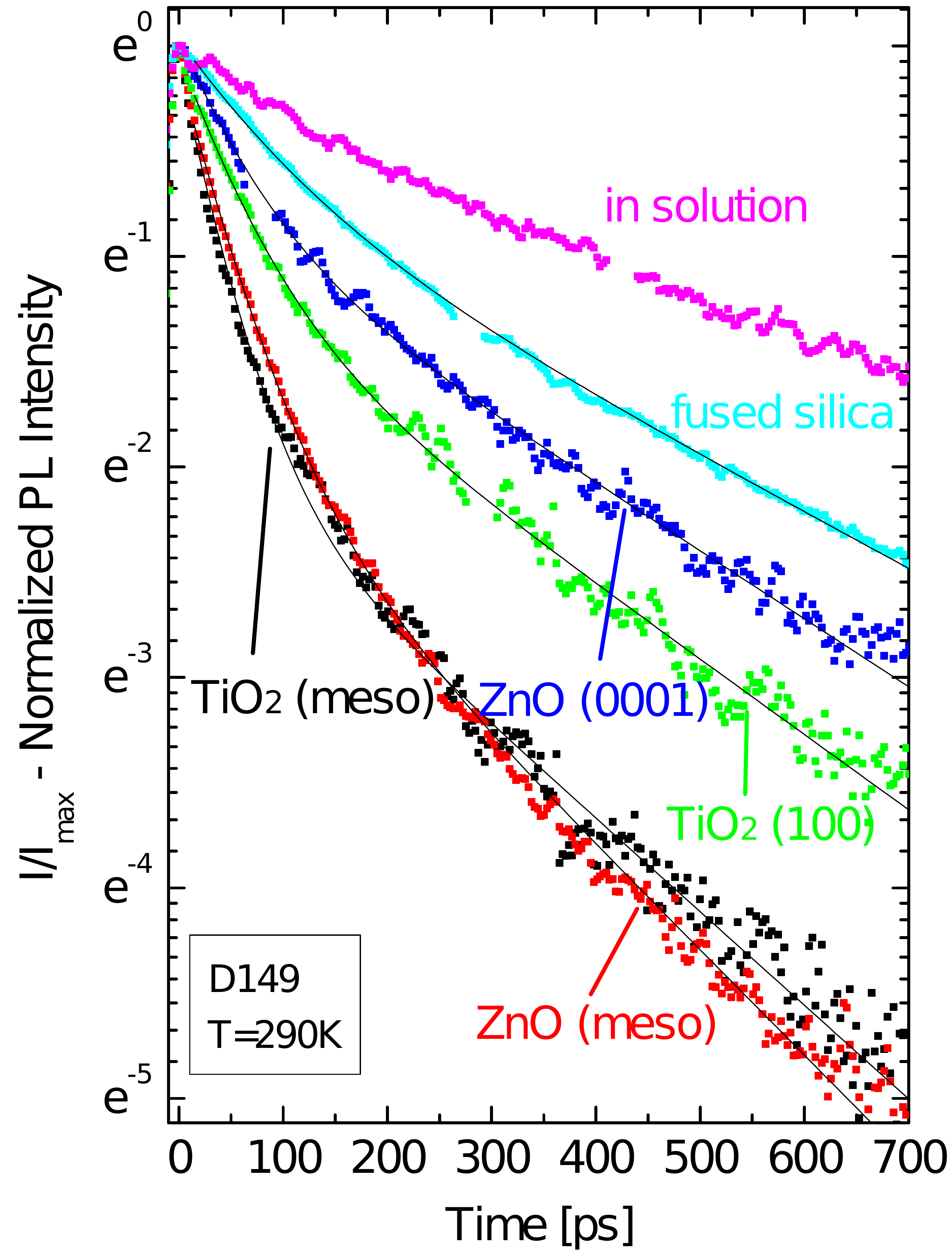}
\caption{Photoluminescence transients of D149 in solution and on various substrates.}
\label{Fig_3}
\end{figure}


In what follows we study the decay dynamics of D149 on ZnO and on TiO$_2$ in detail.
For D149, a conversion efficiency of 6 to 9$\%$ on TiO$_2$(meso) substrates were reported earlier by Horiuchi\cite{Horiuchi2004} and Ito\cite{Ito2006}. The PL decay transients for D149 films on different substrates are summarized in Fig.~\ref{Fig_3}. 
The solution data are given for reference.
Here, the molecules exhibit a long lifetime with an almost exponential decay and a radiative lifetime for the $S_1 \to S_0$ transition of D149 molecules $\tau_{\rm D149}$\,=\,500\,ps.
The residual deviation from the single exponential is tentatively attributed to the so-called concentration quenching effect, i.e., the excitation energy is distributed through radiationless energy transfer between the dye molecules and eventually gets transferred to a nonradiative ''killer'' center. 
Even concentration dependent aggregation of D149 in solution leads to a reduction in lifetime \cite{el2012isomerization}. 
The D149 on both mesoporous ZnO or TiO$_2$ exhibit very fast decays.
Intriguingly, no substantial difference between the mesoporous substrates ZnO and TiO$_2$ is found.
Obviously, additional and much faster decay channels are involved now compared to solution. 
The dominant channel is the transfer of electrons to the interface between the organic layer and the inorganic semiconductor substrate. It should be noted that an alternative loss process, namely the radiationless energy transfer to the inorganic semiconductor, is not possible due to the wide bandgap of the oxide-substrates. 
Contrarily, a thin layer of D149 on a fused silica substrate does not show such a fast decay. 

In general, we ascribe the PL of the dye films to an excitonic recombination with a lifetime $\tau_{\rm exc}$. 
Any additional loss channel changes the respective kinetic description such that it is described according to Eq.~\ref{eq_1} with $w_{\rm trans}=1/\tau_{\rm trans}$ being either the electron transfer probability to the interface or an energy loss by nonradiative centers. Both extra channels reduce the number of radiative excitons. 
Hence, the effective exciton lifetime is then given by Eq.~\ref{eq_2}, which still results in a single exponential decay. 
However, all experimentally observed decay curves are clearly non-single exponential (cf. Fig.~\ref{Fig_3}) inferring a more sophisticated analysis and additional states and channels need to be involved.
It is quite clear that only those excitons close enough to the interface can contribute to the charge transfer. 
Excitons which are too far away from the interface and do not reach the interface during their lifetime cannot dissociate and lose electrons by CT. They will hence recombine with the time constant $\tau_{\rm exc}$. A detailed theoretical description for transients including the mobility of excitons by hopping or diffusion is beyond this paper. 
Hence, we straight-forwardly divide the excitons in those being close enough to the interface and contributing to the CT and those too far away for a CT to take place during their lifetime.

\begin{equation}
\frac{\mathrm d n_{\rm exc}}{\mathrm d t}=-\frac{n_{\rm exc}}{\tau_{\rm exc}}- w_{\rm trans}\cdot n_{\rm exc}
\label{eq_1}
\end{equation}
\begin{equation}
1/\tau_{\rm eff}=1/\tau_{\rm exc} +1/\tau_{\rm trans}
\label{eq_2}
\end{equation} 

This separation yields bi-exponential decay curves given by Eq.~\ref {eq_3}. The weighting factor $n_{\rm a}$ gives the amount of excitons which are created close enough to the interface and can perform either a charge transfer or recombine radiatively. The factor $ n_{\rm b}$ is the number of excitons which are too far away from the interface and will therefore recombine radiatively. It should be noted at this point that the exciton recombination time always includes the radiative as well as possible nonradiative recombination.

Fits of the experimental data using Eq.~\ref {eq_3} are given as black lines in Fig.~\ref{Fig_3}. The exciton lifetimes, the transfer times and the respective weighting factors are given in Table \ref{tab_1}. Note that the electron transfer time from D149 to TiO$_2$ is little shorter compared to ZnO which basically agrees with experimental results of Sobu\'s et al.\cite{Sobus2014}. 
It should be mentioned that the transfer times are longer than the ones reported earlier for D149 on ZnO(meso) which have been determined by transient absorption measurements\cite{Rohwer2015}. This difference is attributed to the fact that we measure the transfer time after the exciton formation and subsequent relaxation into the lowest excited states which allow for a charge transfer. 
It is commonly accepted that the mean transfer time is shorter the higher the electron is excited\cite{Listorti2011, zhang2013comparative}. 

We should mention here other authors using also TRPL to determine charge transfer times reported substantially longer times than ours \cite{cheng2010electron,snaith2009charge}.
Snaith et al. \cite{snaith2009charge} found $\tau_{\rm trans}$\,=\,330\,ps for D149 on TiO$_2$. 
We ascribe such discrepancies to an unavoidable effect caused by different interface to volume ratios. An example of this is evident when comparing the decay curves on mesoporous substrates with the decay curves on flat surfaces (see Fig.~\ref{Fig_3}). The overall slower decay of the films on the smooth substrates is due to the reduced interface area and the respectively higher amount of excitons performing a radiative recombination. This is clearly revealed by the weighting factors (see Table \ref{tab_1}). For the meso-structures we find $n_{\rm a}$ to be about 0.8, i.e. about 80$\%$ of the excitons lose the excited electron by transfer to the substrate, whereas this value is reduced to about 50$\%$ on flat substrates. The relative number of recombining excitons to the number of excitons performing a charge transfer is higher the thicker the organic film. Even the exciton lifetimes of D149 layers (see Table \ref{tab_1}) is different for the meso and flat substrates. We will come back to this point later.

\begin{equation}
n_{\rm exc}(t)=n_{\rm a} \cdot\exp{-\frac{t}{\tau_{\rm eff}}}+n_{\rm b} \cdot\exp{-\frac{t}{\tau_{\rm exc}}}
\label{eq_3}
\end{equation}

We now turn to the decay curve of the D149 thin solid film on fused silica in Fig.~\ref{Fig_3}). It is known that no electron transfer into the silica is possible. Nevertheless, a perfect fit needs again a biexponential function. About 56$\%$ of the excitons recombine radiatively with an excitonic lifetime of about $\tau_{\rm exc}$\,=\,375\,ps. About 44$\%$ of the excitons suffer from another loss channel. We ascribe the loss channel in that case to radiationless transitions, which are always present in the amorphous layers. 

\begin{figure}
\includegraphics[width=0.5\textwidth]{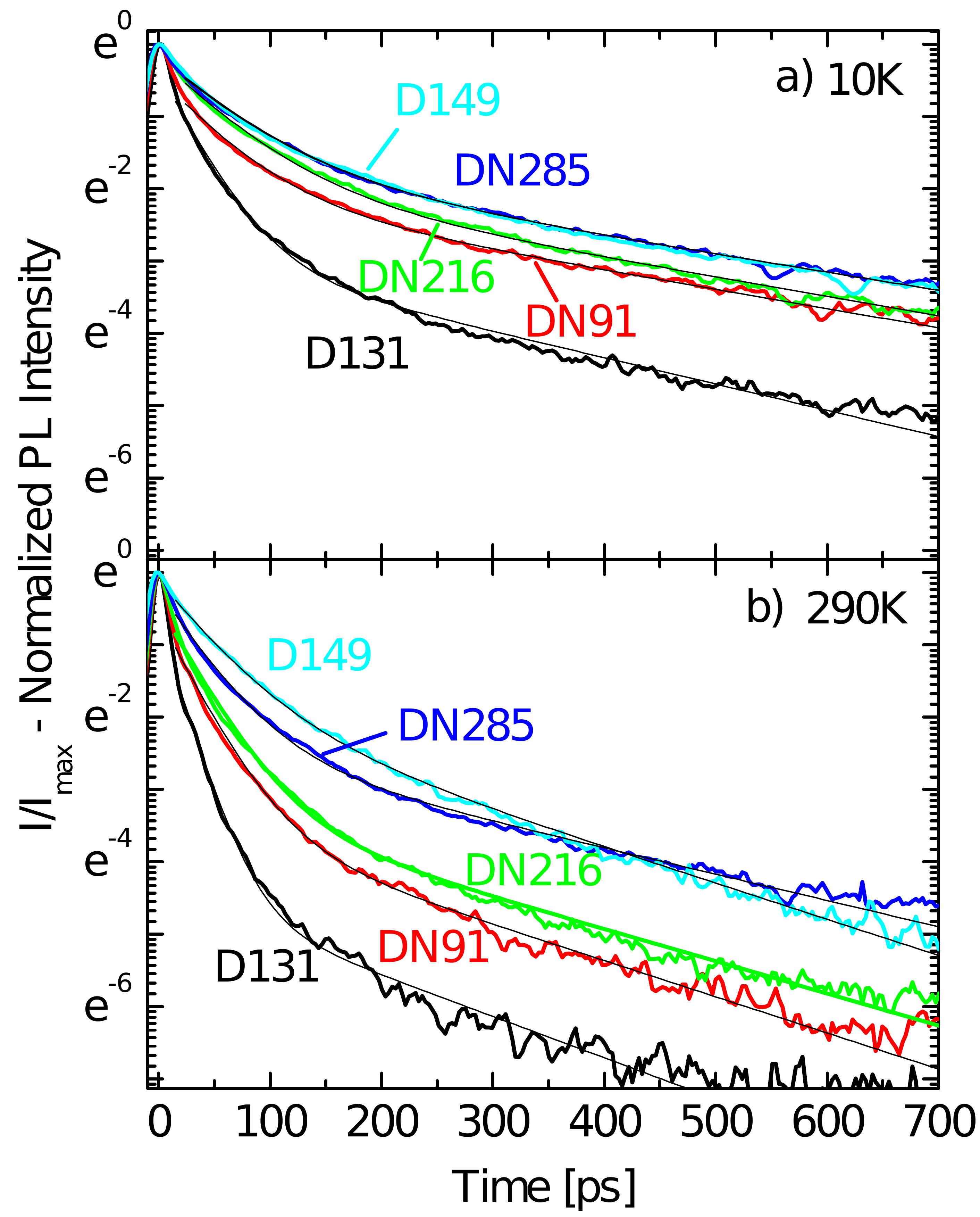}
\caption{Photoluminescence decay of indoline dyes on mesoporous ZnO at temperatures of 10K and 290K.}
\label{Fig_4}
\end{figure}
 
\begin{table}
  \centering
  \caption{Photoluminescence decay times and weighting factors of D149 determined by means of Eq.~\ref {eq_3} (see text for details). The accuracy of all values is $\pm 5\%$.}
    \begin{tabular}{@{}lcccc@{}}
    \toprule
    \textbf{D149 } & $\tau_{\rm trans}$ [ps]  & $n_{\rm a}$ [\%] & $\tau_{\rm exc}$ [ps] & $n_{\rm b}$ [\%] \\
    \colrule
    TiO$_2$(meso) & 46  & 78 & 225     & 22 \\
    ZnO(meso) & 61     & 77 & 200     & 23 \\
    TiO$_2$(100) & 69     & 61 & 280     & 39 \\
    ZnO(0001) & 69     & 51 & 310     & 49 \\
 fused silica  & \textit{115}  & 44 & 375     & 56 \\
    in solution  &      &    & 500  & \\             
    \botrule
    \end{tabular}%
  \label{tab_1}%
\end{table}%

\begin{table}[tb]
  \centering
  \caption{Indoline dyes on mesoporous ZnO at temperatures of 290\,K and 10\,K}
    \begin{tabular}{@{}lcccc@{}}
    \toprule
    \textbf{ZnO (meso)} & $\tau_{\rm trans}$ [ps] & $n_{\rm a}$ [\%] & $\tau_{\rm exc}$ [ps] & $n_{\rm b}$ [\%] \\
    \colrule
    D149 @290K & 61     & 77   & 200   & 23 \\
    DN285 @290K & 47     & 84   & 275   & 16 \\
    DN216 @290K & 41     & 91   & 225   & 9 \\
    DN91 @290K  & 38     & 91   & 200   & 9 \\
    D131 @290K & 23     & 95   & 175   & 5 \\ 
    \colrule  
    D149 @10K & 75     & 71   & 400   & 29 \\
    DN285 @10K & 75     & 71   & 400   & 29 \\
    DN216 @10K & 70     & 76   & 375   & 24 \\
    DN91 @10K & 60     & 73   & 375   & 27 \\
    D131 @10K & 37     & 88   & 275   & 11 \\                    
    \botrule
    \end{tabular}%
  \label{tab_2}%
\end{table}%

\begin{figure}
  \includegraphics[width=0.4\textwidth]{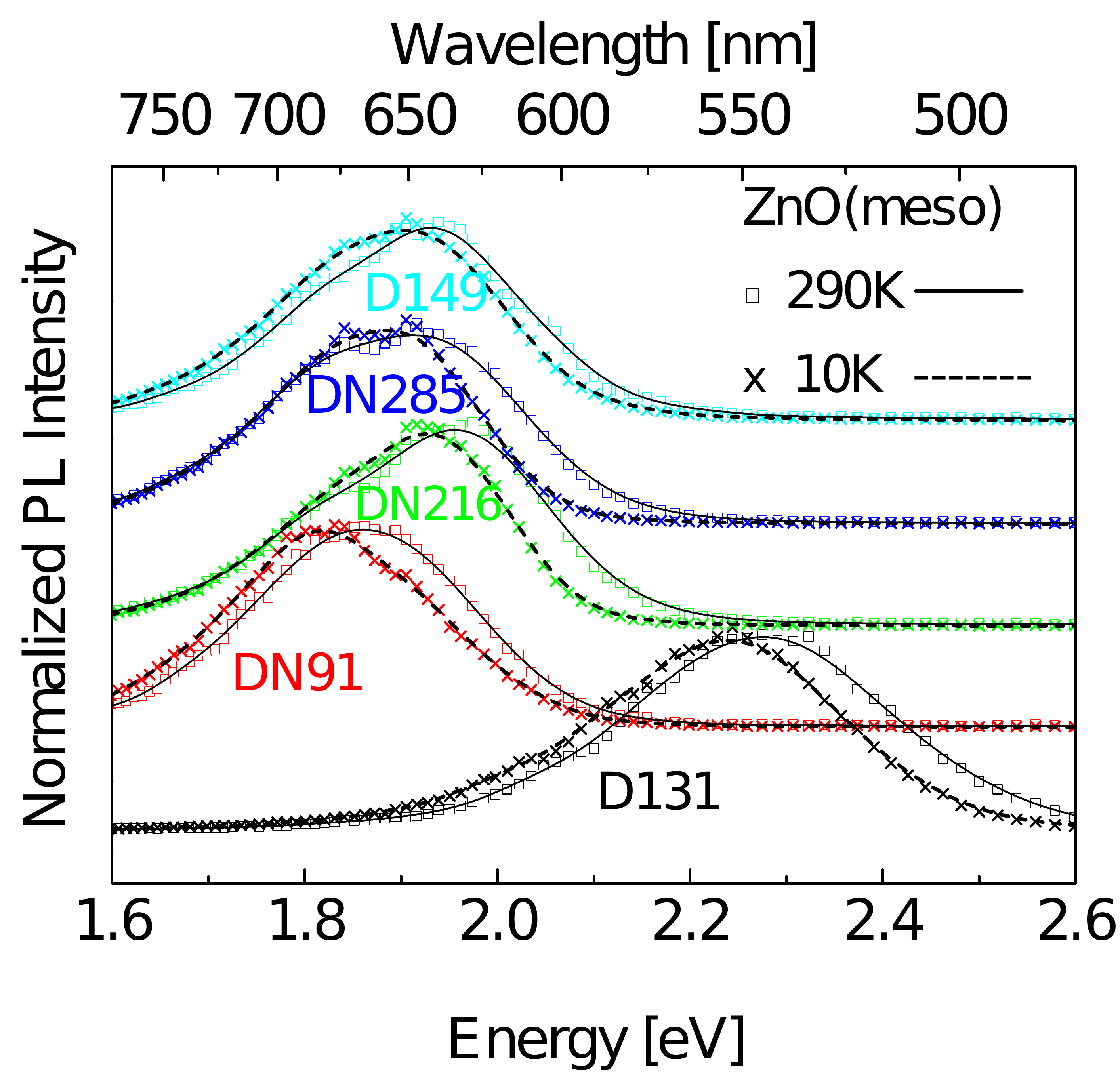}
\caption{Photoluminescence of various dyes on ZnO(meso) at room temperature and T\,=\,10\,K.}
\label{Fig_5}
\end{figure}

       
Next, we study the influence of the anchor groups on the decay dynamics. 
Anchoring groups are expected to improve the energy or charge transfer to the organic-inorganic interface.
The PL transient of the various dyes on ZnO(meso) are depicted in Fig.~\ref{Fig_4}. The results are shown for T\,=\,10\,K (Fig.~\ref{Fig_4}a) and at room temperature (RT) (Fig.~\ref{Fig_4}b). The thin black lines are the best fits using Eq.~\ref {eq_3}. All parameters are given in Table \ref{tab_2}.   

D131 and D149 have both only one carboxyl anchoring group. The charge transfer time is however much faster in case of D131. This is attributed to the fact that due to the blue shift the excited state of D131 is at higher energies above the ZnO-CBM than the corresponding excited state of D149. This interpretation is supported by cyclic voltammetry experiments \cite{fattori2010fast, jose2008conversion} and theoretical calculations \cite{zhang2013comparative, kim2011indoline, howie2008characterization, matsui2011comparison}. A higher energy of the excited state causes a faster charge transfer, in line with earlier findings. To further study this phenomenon we compare the transfer times at RT and T\,=\,10\,K. 
Whereas the band gap of ZnO increases by about 55\,meV by cooling the semiconductor from RT (3,383\,eV \cite{hummer1973interband}) down to T\,=\,10\,K (3,438\,eV at 4\,K \cite{tomm2000optical}) the indoline layers undergo a red shift (see Fig.~\ref{Fig_5}).
The energy difference between the excited electron state and the CBM of ZnO is hence reduced. This leads to a substantially reduced charge transfer probability and respective longer transfer times (see Table \ref{tab_2}).    
We now turn to the second COOH anchor group. Comparing DN91 and D149 reveals that the second anchor improves the CT although the level alignment is expected to be almost identical. Extending the alkyl chain lengths for the second anchor (DN91 $\to$ DN216 $\to$ DN285) increases the CT-time, in good agreement to earlier pump-probe transient absorption measurements at similar samples\cite{Rohwer2015}, and approaches the CT-time of D149 with just one COOH anchor.

\begin{figure}
  \includegraphics[width=0.5\textwidth]{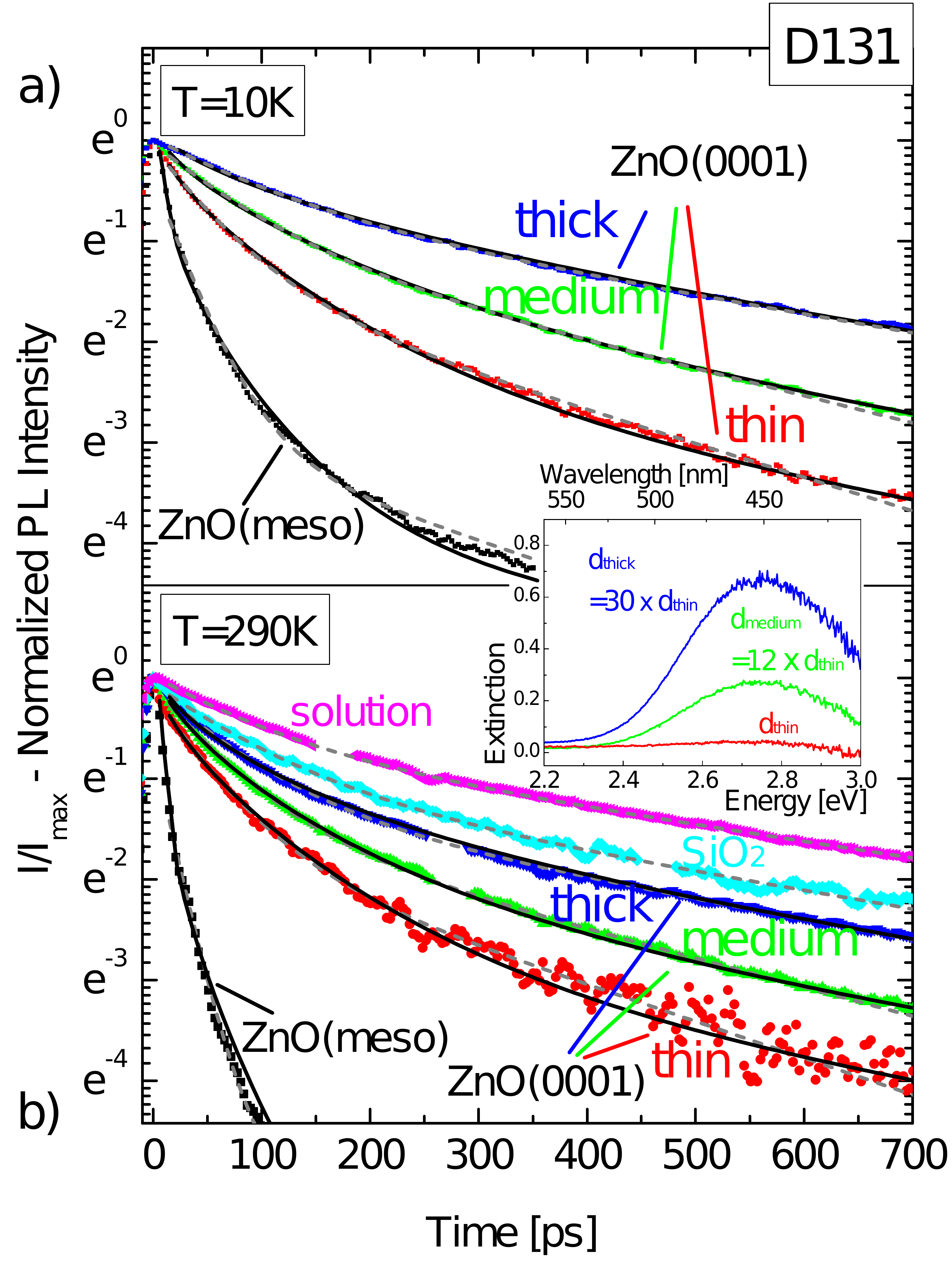}
\caption{Photoluminescence decay of D131 on ZnO. Dashed line: Fit using Eq.3; full lines: Fit using Eq. 6 (see text for details). Inset: Extinction of different layers on ZnO(0001). }
\label{Fig_6}
\end{figure}


\begin{table*}[htb]
  \centering
  \caption{Charge transfer times and exciton recombination times for D131 using Eq.~\ref{eq_3} (upper part) and exciton diffusion time determined with Eq.~\ref{eq_6} (lower part). See text for detailed discussion.}
    \begin{tabular}{lcccc|cccc}
	& \multicolumn{4}{c}{\textbf{T\,=\,290\,K}}  &\multicolumn{4}{c}{\textbf{T\,=\,10\,K}}\\    
    \toprule
    \textbf{D131} \textit{(calc. with eq. 3)} & $\tau_{\rm trans}$ [ps] &  $n_{\rm a}$ [\%] & $\tau_{\rm exc}$ [ps] & $n_{\rm b}$ [\%] 	
    	& $\tau_{\rm trans}$ [ps] &  $n_{\rm a}$ [\%] & $\tau_{\rm exc}$ [ps] & $n_{\rm b}$ [\%]\\
    \colrule   
   in solution  &  &    &  575   &  
   		&	&	&	&\\    
    layer on SiO$_2$(0001)  & \textit{110}   & 59  & 550    & 41 
    	&	&	&	&\\
    	
    thick layer on ZnO(0001) & 90    & 64    & 500   & 36 
    	&  130	&46	&550	&54\\   	
	medium layer on ZnO(0001) & 75    & 68    & 350   & 32 
		&100	&56	&375	&46\\
	thin layer on ZnO(0001)  & 65    & 72    & 275   & 28 
		&  70	&68	&300	&32\\
	ZnO(meso)  & 23    & 95    & 175   & 5 
		& 37    & 88    & 275   & 12\\
    \botrule
    \textbf{D131} \textit{(calc. with eq. 6)} &  $\tau_{\rm D}$[ps] & $n_{\rm a}$ [\%] & $\tau_{\rm exc}$ [ps]  & $n_{\rm b}$ [\%] 
    	&  $\tau_{\rm D}$[ps] & $n_{\rm a}$ [\%] & $\tau_{\rm exc}$ [ps]  & $n_{\rm b}$ [\%]\\
    \colrule  
    
 thick layer on ZnO(0001) &345   & 78    & 550   & 22    & 556   & 63    & 600   & 37 \\
 medium layer on ZnO(0001)&324   & 90    & 550   & 10    & 505   & 85    & 600   & 15 \\
 thin layer on ZnO(0001)  &288   & 95    & 550   & 5     & 334   & 93    & 600   & 7 \\
ZnO(meso)                 &83    & 99    & 550   & 1     & 149   & 97    & 600   & 3 \\
        \botrule
    \end{tabular}%
  \label{tab_3}%
\end{table*}%

Finally, we revisit the aforementioned difference between the transients in case of mesoporous and smooth interfaces. Therefore, we plot the PL transients for D131 on ZnO in Fig.~\ref{Fig_6}. We prepared a series of different layer thicknesses on ZnO(0001). The absorption spectra of the thin, medium, and thick layers are depicted in the inset of Fig.~\ref{Fig_6}. The relative layer thicknesses are determined from the extinction ratio by means of the Beer-Lambert-law (Eq.~\ref{eq_4}) 

\begin{equation}
E_i=lg{\frac{I_0}{I_i}= \epsilon \cdot{d_i}}  
\label{eq_4}
\end{equation}

with $E$ being the extinction, $\epsilon$ the extinction coefficient and $d_i$ ($i$=thick, medium, thin) the respective layer thicknesses. The relative thickness of the dye film on the mesoporous sample could not be evaluated this way due to the much stronger scattering of the porous material, Sakuragi et al. estimated 1.5 monolayer\cite{sakuragi2010aggregation}. The PL transients at T\,=\,10\,K and RT are depicted in Fig.~\ref{Fig_6} a and b, respectively. It can be seen that the overall decay is slower the thicker the film is. This is caused by the increasing amount of excitons which recombine radiatively and do not reach the interface.

Again, we find the slowest decay for a D131 film prepared on silica. Using Eq.~\ref{eq_3} we are able to fit the experimental curve at T\,=\,10\,K as well as at RT, as can be seen by the dashed lines using the parameters given in Table \ref{tab_3} (upper part). 
The times $\tau_{\rm trans}$ and $\tau_{\rm exc}$ increase with increasing layer thickness at RT as well as at 10\,K. Both times are, however, effective parameters as the exciton mobility is not explicitly included in Eq.~\ref{eq_3}. The transfer time $\tau_{\rm trans}$\,=\,23\,ps and $\tau_{\rm trans}$\,=\,37\,ps for D131 on ZnO(meso) at RT and 10\,K are most likely the correct values. For thicker layers $\tau_{\rm trans}$ increases effectively due to the prior exciton diffusion. For the same reason, the effective exciton lifetime gets shorter, the stronger the charge transfer.           
The influence of energy migration on the PL transients, e.g., by hopping or diffusion \cite{faulkner1976effects, yokota1967effects, ghosh1980theory, burshtein1985energy, huber1979fluorescence}, the mechanism of quenching \cite{dexter1953theory}, as well as the dimensionality of the system \cite{chen2007dimensional, berdowski1985energy} remain under discussion since several decades. Regardless, all analytical expressions describe the reality in most cases just approximately. Therefore, numerical procedures are used in some cases \cite{lakowicz1993distance, naumann2000reversible}. 
To take into account the exciton diffusion we modify our model in the following way. The simplest approximation to include a diffusion process is given by the so called Smoluchowski model which assumes an instantaneous quenching of the PL if the fluorophore or excitation energy reaches the quencher by diffusion. By integration of the respective differential equation the intensity decay is given by Eq.~\ref{eq_5} (see e.g.,\cite{lakowicz1993distance}).

\begin{equation}
I(t)=I_0~\text{exp}~[{\frac{-t}{\tau}-2b \sqrt{t}}]  
\label{eq_5}
\end{equation}            

with  $1/\tau=1/\tau_{\rm exc} + \alpha \cdot \rm D$  and b=$\beta \cdot \sqrt{\rm D}$, D is the diffusion parameter. $\alpha$ and $\beta$ are parameters including the interaction radius and molar quencher concentration, which can be considered constant in our case. For the D131 films on ZnO the electron transfer rate is determined by the number of excitons reaching the interface by diffusion. We have to replace the Eq.~\ref{eq_3} by Eq.~\ref{eq_6}.

\begin{equation}
n_{\rm exc}(t)= n_{\rm a} \cdot\exp{-\frac{t}{\tau} -2b \sqrt{t}} +\mathrm n_b \cdot\exp{-\frac{t}{\tau_{\rm exc}}}
\label{eq_6}
\end{equation}

We define a mean diffusion time $\tau_{\rm D} =1/{(\alpha \cdot \rm D)}$ to eliminate the unknown diffusion parameter, yielding $1/\tau=1/\tau_{\rm exc} +1/\tau_{\rm D}$ and b=$\beta^\prime \cdot \sqrt{1/\tau_{\rm D}}$.     
Applying Eq.~\ref{eq_6} to the decay curves in Fig.~\ref{Fig_6} yields a perfect fit (full black lines). As already mentioned, the overall decay becomes slower with increasing layer thickness. This is attributed to the lengthy diffusion process of excitons towards the interface, which is given in the model by a diffusion time. As can be seen in Table \ref{tab_3} (lower part) the exciton lifetime $\tau_{\rm exc}$\,=\,550\,ps at RT is now the same for all samples but the mean diffusion time increases from 83\,ps for D131 on ZnO(meso) to about 345\,ps for the thickest layer. It is particularly interesting to note that almost all the excitons suffer an electron loss by charge transfer in case of the ZnO(meso) structure. The value is reduced to about 78\% for the thick layer at RT. 
The exciton lifetime at T$ = 10\,$K of the D131 layers $\tau_{\rm exc}$\,=\,600\,ps is longer than the value at RT due to reduced radiationless losses. The longer lifetime should basically enhance the probability for charge transfer. As can be seen in table \ref{tab_2} and also in the upper part of table \ref{tab_3}, the charge transfer is reduced in all cases. 
The explanation for this surprising behavior is now given in table \ref{tab_3} (lower part). The diffusion times are substantially longer at T\,=\,10\,K compared to RT, which obviously hints at a strongly reduced mobility at T\,=\,10\,K. This clearly reveals exciton hopping as the underlying mobility process.

\begin{figure}
  \includegraphics[width=0.5\textwidth]{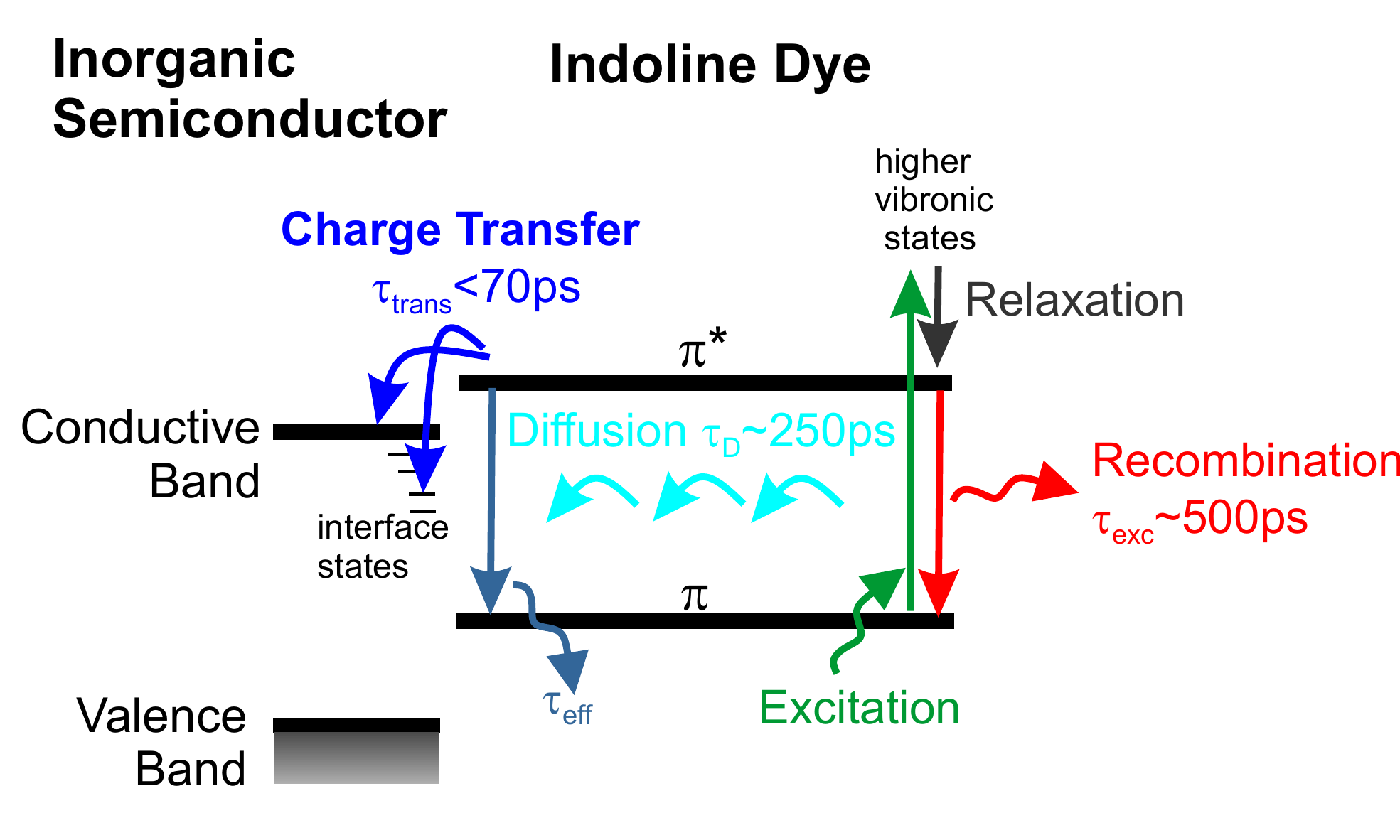} 
\caption{Schematic picture of the processes at the interface between the indoline layer and inorganic semiconductor.}
\label{Fig_7}
\end{figure}


\section{Conclusion}
\label{sec:conclusion}

In conclusion, we studied the charge transfer from excitons in indoline dye layers to the organic-inorganic interface. The external quantum efficiency of solar cells is the most important parameter for practical use. It is defined by the number of electrons flowing trough the external circuit divided by the number of incident photons. The charge transfer, though very important, is only one step in others and used to be not known. We present here a method to determine independently the charge transfer time for dye semiconductor hybrids. The relevant processes, namely excitation, recombination, diffusion, and charge transfer are depicted schematically in Fig.7.
Comparing D149 and DN91 reveals that the second COOH-anchor and expected enhanced bonding strength improves the charge transfer. Enhancing the distance between the chromophore and the interface by extending the anchor chain (DN91 $\to$ DN216 $\to$ DN285) reduces the transfer probability. The fastest transfer was found for the D131. This is ascribed to larger energy difference of the excited dye state and the conduction band minimum of the bulk semiconductor. The relevance of the level alignment is supported by the temperature dependence of the charge transfer manifested in the PL decay times: the transfer probability into the ZnO is reduced with lower temperature for all dyes. The reason is the increasing band gap of the semiconductor but decreasing $\pi\to\pi^\star $ transition energies. This results in a reduced energetic difference between the excitonic states of the dyes with the semiconductor CBM.\\ 
Finally, we revealed the influence of the excitonic mobility on the charge transfer. Increasing dye layer thicknesses enhances the absorption and hence the total number of charge transfer electrons, but reduces the quantum efficiency due to the increasing number of excitons which do not reach the interface and cannot contribute to the charge transfer.


\begin{acknowledgments}
We are grateful to financial support by the DFG in the framework of the SFB 1083 (IM, NWR, SC and WH) and within project SCHL 340/19-1 (JF and DS). 
\end{acknowledgments}

\bibliographystyle{charge_transfer_bibl} 
\bibliography{charge_transfer_bibl}

\end{document}